\documentclass[prc,superscriptaddress,nofootinbib,twocolumn,floatfix]{revtex4}
\usepackage{amsfonts}
\usepackage{graphicx,color,amsmath,amssymb}
\usepackage{subcaption}

\def\blfootnote{\xdef\@thefnmark{}\@footnotetext}
\begin{document}

\title{Laboratory-frame $T$-matrix and heavy quark drag in the quark-gluon plasma}

\author{Anurag Tiwari$^{1}$ and Min He$^{1}$}
\affiliation{$^1$Department of Applied Physics, Nanjing University of Science and Technology, Nanjing~210094, China}

\date{\today}

\begin{abstract}
Non-perturbative scattering $T$-matrix is a core input for the evaluation of transport phenomena in a strongly-coupled medium. Existing in-medium $T$-matrix calculations are typically formulated in the two-particle center-of-mass frame, where the scattering equation can be reduced to a lower-dimensional problem. However, a medium explicitly breaks Lorentz invariance and defines a preferred reference frame, entailing that physical observables be constructed from scattering amplitudes evaluated in the medium rest (laboratory) frame. In this work, by exploiting the rotational symmetry about the scattering-pair-momentum axis, we develop a practical framework for solving the in-medium two-body $T$-matrix directly in the laboratory frame while retaining the full dependence on the total pair-momentum and scattering geometry. We demonstrate that the resulting amplitudes differ significantly from conventional center-of-mass-frame results and, when applied to heavy-light quark scattering in the quark-gluon plasma (QGP), lead to 25-40\% corrections to heavy-quark drag coefficients at low momenta, thereby removing a significant source of theoretical uncertainty in extracting the QGP transport properties with heavy-quark probes.
\end{abstract}

%\pacs{25.75.Dw, 12.38.Mh, 25.75.Nq}

\maketitle

%%%%%%%%%%%%%%%%%%%%%%%%%%%%%%%%%%%%%%%%%%%%%%%%%%%%%%%%
\section{Introduction}
\label{sec_intro}
%%%%%%%%%%%%%%%%%%%%%%%%%%%%%%%%%%%%%%%%%%%%%%%%%%%%%%%%
Non-perturbative two-body scattering amplitudes, commonly referred to as $T$-matrices, constitute a fundamental building block in the theoretical description of transport phenomena across different physical contexts, ranging from heavy quark diffusion in the quark-gluon plasma (QGP)~\cite{Rapp:2009my,Rapp:2018qla,Dong:2019unq,He:2022ywp}, nuclear reactions in finite nuclei and bulk nuclear matter~\cite{TerHaar:1986xpv,Dickhoff:2004xx,Colonna:2020euy} to the Sommerfeld-enhanced annihilation of Dark Matter particles in the heat bath of the early Universe~\cite{Hisano:2004ds,Arkani-Hamed:2008hhe,Kim:2016kxt,Qerimi:2025aga}. The accurate determination of these in-medium amplitudes is therefore essential for precision extraction of the underlying medium properties.

The computation of such in-medium T-matrices typically proceeds by solving the Lippmann-Schwinger-type integral equation in the two-particle center-of-mass (CM) frame~\cite{Dickhoff:1999yi,Dickhoff:2004xx,Hisano:2004ds,Mannarelli:2005pz,Riek:2010fk,Liu:2017qah}. In this frame, the T-matrix reduces to a lower-dimensional quantity owing to the intact spherical symmetry and thereby the scattering equation can be simplified by partial-wave expansion techniques~\cite{Haftel:1970zz}, allowing for efficient numerical solution. The scattering amplitudes thus obtained are then used as inputs for evaluating medium transport properties~\cite{Dickhoff:1999yi,Dickhoff:2004xx,Hisano:2004ds,Dickhoff:1998zz,vanHees:2007me,Riek:2010fk,Liu:2018syc,Tang:2023tkm}.

However, the three-dimensional formulation of the Lippmann-Schwinger-type scattering equation utilizing an instantaneous potential already explicitly breaks the Lorentz invariance even in vacuum. In a medium, this is further compounded by the fact that the medium defines a preferred reference frame - the medium rest (or laboratory) frame. Consequently, the amplitudes in the CM- and laboratory-frame are not simply related by a Lorentz boost and physical observables must be constructed from scattering amplitudes evaluated directly in the laboratory frame. Technically, the presence of a finite total pair-momentum in the laboratory frame breaks the spherical symmetry of the scattering geometry, making it impossible to reduce the scattering equation into independent partial-wave channels~\cite{Schiller:1998ff,Frick:2003sd,Soma:2006zx}. 

The assumption that the CM-frame amplitude adequately approximates the laboratory-frame amplitude for in-medium transport calculations is simply based on the implicit belief that the dependence on the total pair momentum is weak enough to be negligible~\cite{Dickhoff:1999yi,Liu:2017qah}, but has not been subjected to a systematic quantitative test. %to the best of our knowledge.
The heavy-quark diffusion in the strongly coupled QGP provides a compelling case for addressing this frame mismatch. In particular at low momentum, the large mass of heavy quarks ({\it e.g.} $m_Q\simeq 1.5$\,GeV for charm) entails that they thermalize via multiple elastic collisions with QGP particles with relatively small momentum transfer of the order of temperature of the medium, $q\sim T$. This enables their propagation in the QGP to be treated as Brownian motion and as such described by a stochastic Langevin process characterized by a transport parameter -- the heavy quark drag coefficient~\cite{Svetitsky:1987gq,Moore:2004tg,Rapp:2009my,Sambataro:2024mkr,Altenkort:2023oms,HotQCD:2025fbd}, providing an excellent window into the inner workings of the near-ideal QGP liquid~\cite{Dong:2019unq,He:2022ywp,Xu:2017obm,Dong:2019byy,He:2019vgs,Beraudo:2022dpz,Liu:2021dpm,Zhao:2023nrz,Pandey:2023dzz,Krishna:2025bll,Sambataro:2025pop}. Furthermore, the energy transfer $q_0\simeq q^2/m_Q$ ($m_Q$$\gg$$T$) in these elastic collisions is parametrically suppressed relative to the momentum transfer, justifying the potential-type interaction which can be conveniently resummed and unitarized by the $T$-matrix equation to account for the strong coupling of heavy quarks with the medium~\cite{Rapp:2009my,Andronic:2017pug,Liu:2018syc,Dong:2019byy,He:2019vgs,Brambilla:2020siz,Altenkort:2023oms}, as primarily evidenced by the large elliptic flows of charm-hadrons~\cite{ALICE:2021rxa,ALICE:2026zcz,CMS:2017vhp,STAR:2017kkh,ALICE:2017quq,ALICE:2020pvw,He:2021zej}. 

Indeed, over the past two decades a systematic $T$-matrix approach has been developed to address the heavy quark non-perturbative interactions in the QGP~\cite{vanHees:2007me,Riek:2010fk,Liu:2017qah,Tang:2023tkm}, yielding a heavy quark drag coefficient close to the value determined by lattice QCD~\cite{Altenkort:2023oms,HotQCD:2025fbd, Brambilla:2020siz,Altenkort:2020fgs,Banerjee:2022gen,Altenkort:2023eav} at vanishing momentum and admitting a marked enhancement relative to perturbative calculations~\cite{Moore:2004tg,Caron-Huot:2007rwy}. Despite their phenomenological success, these $T$-matrix calculations have so far been restricted to the two-particle CM frame, leaving the impact of the frame mismatch on heavy-quark transport coefficients entirely unexplored. Given the critical role of heavy quarks in probing QGP properties in the precision measurement era~\cite{ALICE:2022wpn,ALICE:2022wwr,CMS:2024irj}, quantifying the uncertainty arising from this mismatch now represents a theoretical priority.

In this work, we develop a practical framework to solve the in-medium $T$-matrix as a five-dimensional quantity directly in the laboratory frame, retaining the full dependence on the total pair-momentum and scattering geometry. Exploiting the azimuthal symmetry about the pair-momentum axis as a remnant of the breaking of spherical symmetry for the {\it scattering state}, we show that the scattering equation for the laboratory-frame $T$-matrix can be decomposed into independent, uncoupled integral equations for the azimuthal components, $T_m$, allowing for numerically tractable implementations. We demonstrate that the full laboratory-frame scattering amplitudes differ significantly from the CM-frame counterparts. When applied to heavy-light quark scattering in the QGP, this exact treatment yields a 25-40\% correction to the heavy-quark drag coefficient at low momenta. 

%%%%%%%%%%%%%%%%%%%%%%%%%%%%%%%%%%%%%%%%%%%%%%%%%%%%%%%%%%%%%%%%
\section{Laboratory-frame $T$-matrix equation}
\label{sec_lab-frame-T-matrix}
%%%%%%%%%%%%%%%%%%%%%%%%%%%%%%%%%%%%%%%%%%%%%%%%%%%%%%%%%%%%%%%%
%With single-particle states non-relativistically normalized $<\vec q'|\vec q>=(2\pi)^3\delta^3(\vec q'-\vec q)$
The $T$-matrix equation for two-particle scattering in a homogeneous medium through a non-relativistic, instantaneous potential $V$ is given by~\cite{Kadanoff-Barym,Danielewicz:1982kk,Botermans:1990qi}
\begin{align} \label{T_matrix_general}
	T(E, \vec{P}, \vec{q'}, \vec{q}) &= V(\vec{q'}, \vec{q}) + \int \frac{d^{3} k}{(2\pi)^3}V(\vec{q'} , \vec{k}) G(E, \vec{P},\vec{k}) \nonumber\\
	&\times T(E, \vec{P}, \vec{k}, \vec{q})
\end{align}
where $E$, $\vec{P}$ represent the total energy and total (center-of-mass) momentum of the pair, respectively; $\vec q$ and $\vec q'$ denote the relative momentum in the initial and final state. Including both particle-particle and hole-hole channels, the non-interacting two-particle propagator for dressed particles in Eq.~(\ref{T_matrix_general}) reads
\begin{align}
	\label{eq: two_body_prop_SF_form}
	G(E, \vec{P},\vec{k}) &= \int_{-\infty}^{\infty}\frac{d\omega}{2\pi}\int_{-\infty}^{\infty}\frac{d\omega'}{2\pi} \frac{1 - n(\omega) - n(\omega')}{E - \omega -\omega'+i\epsilon} \nonumber \\
	&\times\rho_{1}(\omega, |\frac{\vec{P}}{2}+\vec{k}|)\rho_{2}(\omega', |\frac{\vec{P}}{2}-\vec{k}|),
\end{align}
Medium effects in this formalism are thus encoded in the potential $V$, the Bose/Fermi distribution functions $n(\omega)$ and the single-particle spectral functions $\rho(\omega,\vec p)$.

Let's first recall the conventional partial wave $T$-matrix equation in the CM frame. We consider a local, central potential that admits a Fourier transform
\begin{align}
	\label{eq: potential_Fourier_transorm}
	V(\vec{q}',\vec{q}) = \int d^3r\;e^{i(\vec{q}-\vec{q'})\cdot\vec{r}} V(r).
\end{align}
The $SO(3)$ symmetry entails $X(\vec{q'},\vec{q}) = X(q',q,\hat{\vec q'}\cdot\hat{\vec q})$ and thus the partial wave expansion
\begin{align}\label{CM-PWE}
	%X(\vec{q'},\vec{q}) = 4\pi\sum_{l=0}^{\infty}(2l+1)X_l(q',q)P_l(\rm cos \theta_{q'q}),
	X(\vec{q'},\vec{q}) = 4\pi\sum_l(2l+1)X_l(q',q)P_l(\rm cos \theta_{q'q}).
\end{align}
Substituting this into Eq.~(\ref{T_matrix_general}) for $X=T$ or $V$ with $\vec P=0$, and noting that the two-particle propagator $G(E_{\rm CM},k)$ depends only on $k=|\vec k|$ of the relative momentum in the intermediate state due to spherical symmetry, the integration over the angular direction of $\vec k$'s can be performed using the orthogonality relation of the Legendre polynomials
\begin{align}\label{orthogonality-Legendre-polynomials}
	\int d\Omega_k P_l(\hat{\vec q'}\cdot\hat{\vec k})P_{l'}(\hat{\vec k}\cdot\hat{\vec q})=\frac{4\pi}{2l+1}\delta_{ll'}P_l(\hat{\vec q'}\cdot\hat{\vec q}),
\end{align}
yielding the partial wave $T$-matrix equations
\begin{align}
	\label{eq: partial-wave-T-matrix}
	T_l(E_{\rm CM},q',q) &= V_l(q',q) + \frac{2}{\pi}\int k^2dk V_l(q',k) G(E_{\rm CM}, k) \nonumber\\
	&\times T_l(E_{\rm CM},k,q).
\end{align}

Now moving to the laboratory frame, the $SO(3)$ symmetry dictates that the $T$-matrix remains a function of rotational scalars $T=T(E,P,q',q,\hat{\vec P}\cdot\hat{\vec q'},\hat{\vec P}\cdot\hat{\vec q},\hat{\vec q'}\cdot\hat{\vec q})$. But the pair-momentum $\vec P$ defines a preferred direction that breaks the $SO(3)$ down to a smaller $SO(2)$ rotational symmetry about the $\vec P$-axis for the {\it scattering state}. Indeed by aligning $\vec P$ along the $z$-axis, the dependence of $T$ or $V$ on the azimuthal angles $\phi_q$ and $\phi_{q'}$ enters only through their difference via $\hat{\vec q'}\cdot\hat{\vec q}={\rm cos}\theta_{q'}{\rm cos}\theta_{q}+{\rm sin}\theta_{q'}{\rm sin}\theta_{q}{\rm cos}(\phi_{q'}-\phi_{q})$, allowing for an expansion in azimuthal harmonics analogous to Eq.~(\ref{CM-PWE})
\begin{align}\label{Lab-azimutal-expansion}
	%X(\phi_{q'}-\phi_q) = \sum_{m=\infty}^{\infty} X_m e^{im(\phi_{q'}-\phi_q)},
	X(\phi_{q'}-\phi_q) = \sum_m X_m e^{im(\phi_{q'}-\phi_q)},
\end{align}
for $X=T$ or $V$ (with dependence on other variables suppressed). Substituting this into Eq.~(\ref{T_matrix_general}) and observing that the two-particle propagator $G(E,P,k,{\rm cos}\theta_k)$ also respects the azimuthal symmetry (i.e. independent of $\phi_k$~\cite{Holz:1988pm,Elster:1997hp,Caia:2003ke}), the integration over $\phi_k$ can be carried out analytically using the orthogonality relation
\begin{equation}
	\int_{0}^{2\pi}d\phi_k e^{i m (\phi_{q'}-\phi_k)}e^{i m' (\phi_{k}-\phi_q)} = 2\pi\delta_{mm'}e^{i m (\phi_{q'}-\phi_q)},
\end{equation}
yielding the azimuthal $T$-matrix equations
\begin{align}
	\label{T_m-equations}
	T_m(E,P,q',q,x',x) = V_m(q',q,x',x) +  \int\frac{k^2dk}{(2\pi)^2} dx''  \nonumber\\
	\times V_m(q',k,x',x'')G(E,P,k,x'')T_m(E,P,k,q,x'',x),
\end{align}
where $x=\hat{\vec P}\cdot\hat{\vec q},x'=\hat{\vec P}\cdot\hat{\vec q'}$ and $x''=\hat{\vec P}\cdot\hat{\vec k}$.  These integral equations, like Eq.~(\ref{eq: partial-wave-T-matrix}), are uncoupled from each other as guaranteed by the azimuthal symmetry. Conversely, projection onto a conventional partial-wave ($l$) basis would result in a highly coupled, computationally demanding system~\cite{Cheon:1988hn,Schiller:1998ff} due to the breaking of spherical symmetry in the two-particle propagator at finite $\vec P$.

One notes that for a real potential, $V_m=V_{-m}$ and $T_m=T_{-m}$, so that the full amplitude is constructed as
\begin{align}\label{Lab-full-T-matrix}
	T(\Phi) = T_0 + \sum_{m=1}^{\infty} 2T_m {\rm cos}(m\Phi).
\end{align}
with $\Phi=\phi_{q'}-\phi_q$. In the $\vec P=0$ limit, the spherical symmetry is restored and only the isotropic $m=0$ component $T_0$ survives, recovering the CM frame $T$-matrix equation without partial wave expansion~\cite{Elster:1997hp,Tiwari:2025vju}.
%[since then one can align $\vec q$ with the $z$-axis and restrict $\vec q'$ within the $xz$-plane such that $\phi_q=\phi_{q'}=0$ without loss of generality].

%%%%%%%%%%%%%%%%%%%%%%%%%%%%%%%%%%%%%%%%%%%%%%%%%%%%%%%%%%%%%%%%
\section{Laboratory- vs CM-frame $T$-matrix}
\label{sec_lab-vs-cm-T-matrix}
%%%%%%%%%%%%%%%%%%%%%%%%%%%%%%%%%%%%%%%%%%%%%%%%%%%%%%%%%%%%%%%%
Having established the uncoupled azimuthal component framework, we now embark on numerical implementations of the laboratory-frame $T$-matrix and contrast them with CM-frame calculations. For a concrete application, we examine the scattering between a heavy (charm) quark of mass $m_Q = 1.3\text{ GeV}$) and a light antiquark of thermal mass $m_q = 0.4\text{ GeV}$ in a QGP at temperature $T = 0.19\text{ GeV}$ with a screening mass $\mu=0.25$\,GeV, interacting via a screened Cornell potential $V(r, T) = -\alpha e^{-\mu r}/r - \sigma e^{- \mu r}/\mu$~\cite{Karsch:1987pv} (with the infinite-distance term subtracted and taken as a mean-field contribution to the heavy quark mass~\cite{Riek:2010fk}), which admits an analytical Fourier transform
\begin{align}\label{momentum-space-potential}
	V(\vec{q'}, \vec{q}) & = -\frac{4\pi \alpha}{(\vec{q'}- \vec{q})^2 + \mu^2} - \frac{8\pi \sigma}{((\vec{q'}- \vec{q})^2 + \mu^2)^2}.  %\nonumber \\
	%(\vec{q} - \vec{q'})^2 & = q^2 +q'^2-2q q'(\sqrt{1-x^{'2}}\sqrt{1-x^2}\cos{\Phi} + x' x),
\end{align}
The quark spectral function $\rho(\omega, \vec{p}) = \Gamma(\omega,\vec{p})/[(\omega - \omega_{p})^2 + \Gamma(\omega,\vec{p})^2/4]$ (neglecting the real part of self-energies other than the aforementioned mean-field contribution) in Eq.~(\ref{T_matrix_general}) in principle should be determined together with the $T$-matrix in a self-consistent way~\cite{Liu:2017qah}. Refraining from performing such a calculation which is beyond the purpose of the present work, we instead employ a constant width and bracket the pertinent uncertainty by varying $\Gamma=100$-$200$\,MeV~\cite{Riek:2010fk}.

Subjected to two on-shell conditions (identically for $\vec{q'}$)
\begin{align}\label{on-shell-conditions}
	E=\sqrt{(\vec P/2+\vec q)^2+m_Q^2}+\sqrt{(\vec P/2-\vec q)^2+m_q^2},
\end{align}
the amplitude $T_{\rm Lab}(E,P,x,x',\Phi)$ is a function of five variables that together dictate a full scattering configuration in the laboratory frame. Conversely, the on-shell CM-frame amplitude $T_{\rm CM}(q_{\rm CM},\theta_{\rm CM})$ depends only on two variables. To ensure that both $T_{\rm Lab}$ and $T_{\rm CM}$ evaluate the same physical scattering event, the momenta of the incoming and outgoing particles in the laboratory frame are first determined from $(E,P,x,x',\Phi)$. These momenta are then Lorentz transformed to obtain the relative momentum $q_{\rm CM}$ and the scattering angle $\theta_{\rm CM}$ in the CM frame using $\vec \beta=\vec P/E$ and $\gamma=1/\sqrt{1-\beta^2}$.

\begin{figure*}[t]   
	\centering
	\begin{subfigure}{0.24\textwidth}
		\centering
		\includegraphics[width=\linewidth]{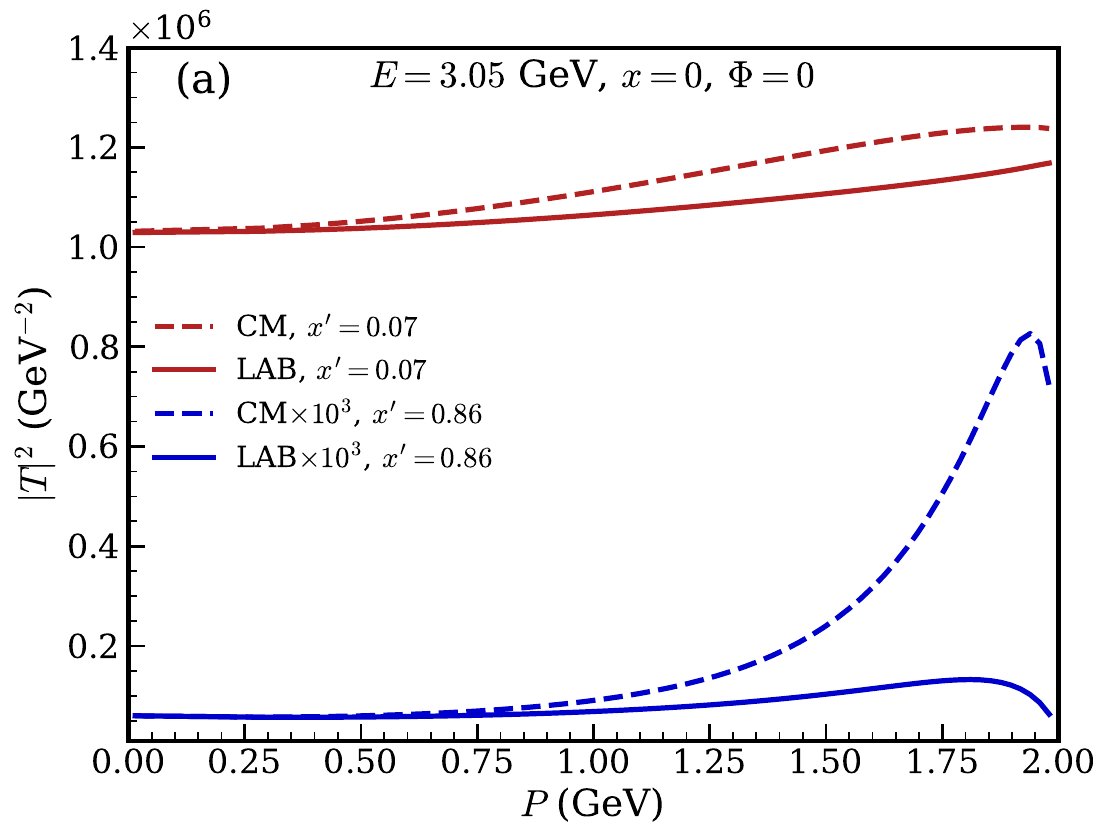}
		%\caption{1}
		\label{fig:sub1}
	\end{subfigure}
	\hfill
	\begin{subfigure}{0.24\textwidth}
		\centering
		\includegraphics[width=\linewidth]{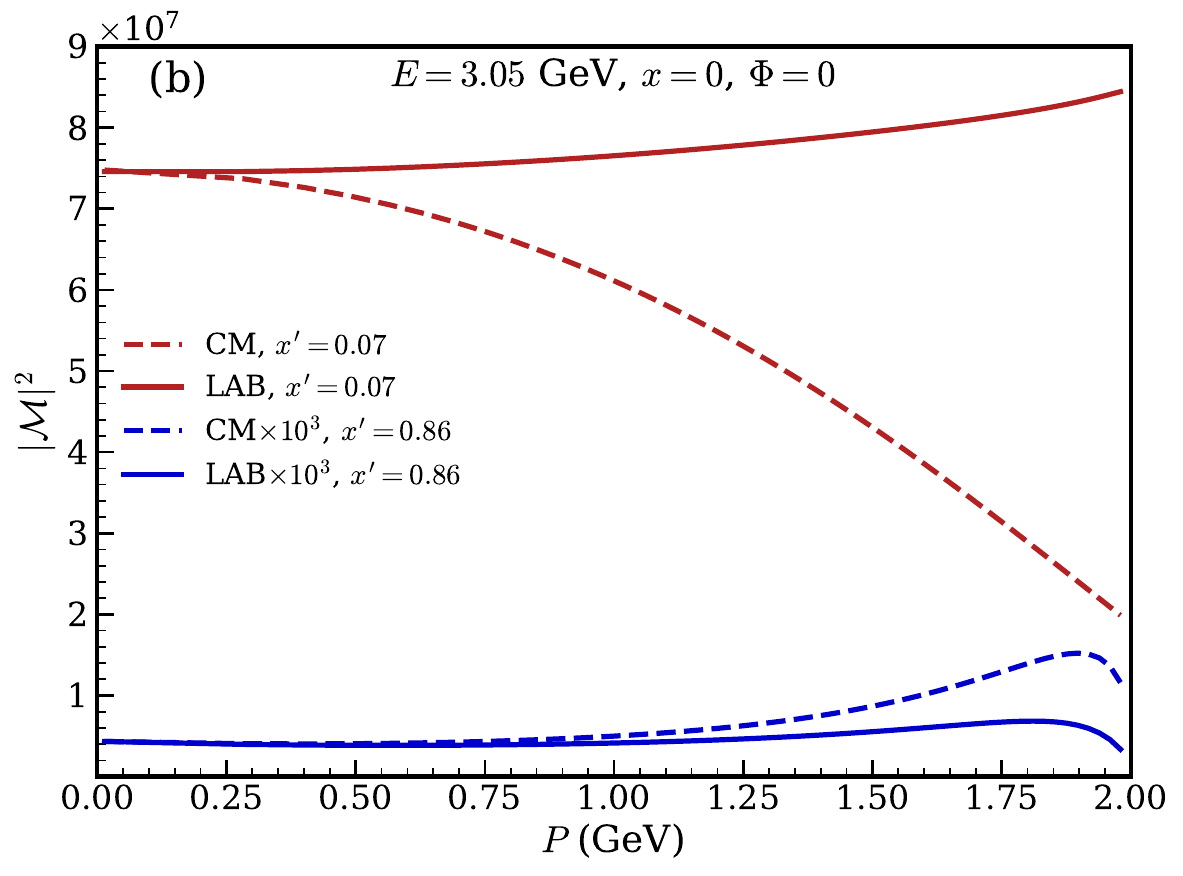}
		%\caption{2?}
		\label{fig:sub2}
	\end{subfigure}
	\hfill
	\begin{subfigure}{0.24\textwidth}
		\centering
		\includegraphics[width=\linewidth]{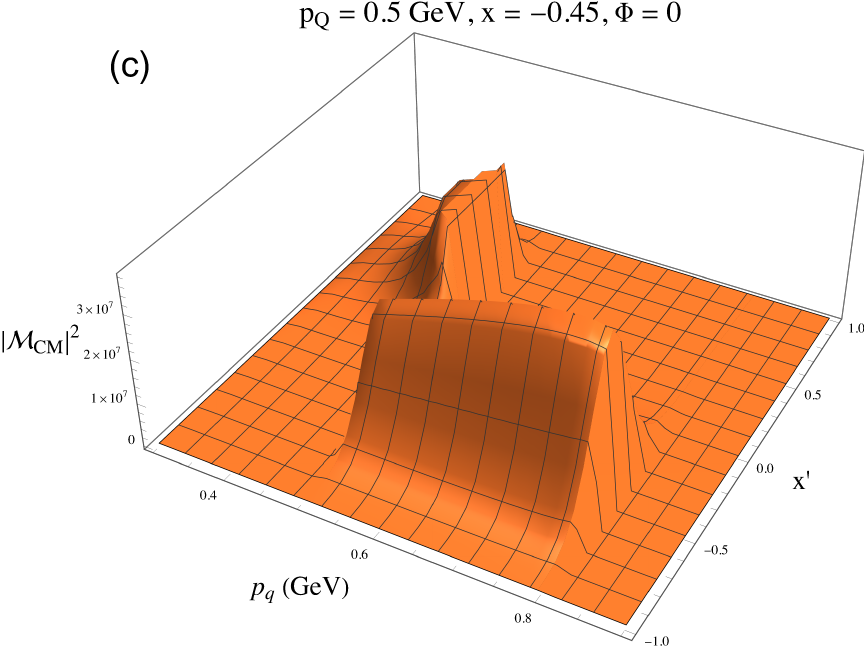}
		%\caption{3}
		\label{fig:sub3}
	\end{subfigure}
	\hfill
	\begin{subfigure}{0.24\textwidth}
		\centering
		\includegraphics[width=\linewidth]{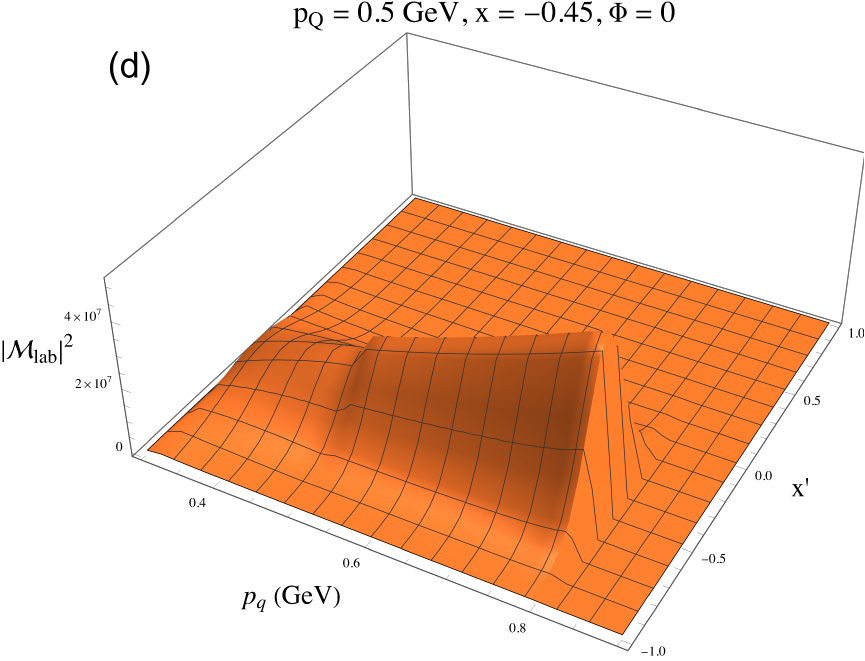}
		%\caption{4}
		\label{fig:sub4}
	\end{subfigure}
	\caption{(Left two) The laboratory- {\it v.s.} CM-frame comparison for the charm-light quark scattering $|T|^2$ and the relativistically normalized $|\mathcal{M}|^2$ as a function of the pair-momentum $P$ at $T$=190\,MeV in the QGP. (Right two) The same comparison for  $|\mathcal{M}|^2$ projected onto the two-dimensional plane of laboratory-frame variables ($p_q,x'$).}
	\label{fig:Lab-vs-CM-amplitudes-comparison}
\end{figure*}

To quantify the difference between the laboratory- and CM-frame amplitudes, we consider the following scattering configuration. We fix the incoming quark momenta $p_Q=p_q=1$\,GeV so that the total energy $E=\omega_Q+\omega_q$ is fixed and the total momentum %$P=\sqrt{p_Q^2+p_q^2+2p_Qp_qx_{Qq}}$ changes in $[0,2]$\,GeV by varying the angle $x_{Qq}=\hat{\vec p}_Q\cdot\hat{\vec p}_q$
$P$ changes within $[0,2]$\,GeV, while the cosine of the angle between $\vec P$ and $\vec q$ identically vanishes ($x=0$). We set $\Phi=0$, which implies all momentum vectors are within the same plane and all $T_m$ components add up constructively in Eq.~(\ref{Lab-full-T-matrix}). By varying the outgoing angle $x'$, the forward ($x'\approx x$) and non-forward ($x'$ differing significantly from $x$) regions are probed. As demonstrated in Fig.~\ref{fig:Lab-vs-CM-amplitudes-comparison}(a), at $P=0$ the amplitudes computed in two frames coincide ($|T_{\text{Lab}}|^2 = |T_{\text{CM}}|^2$), indicating that our numerical implementation correctly reproduces the CM-frame result in the appropriate limit. However, as $P$ increases, a distinct divergence between these two amplitudes emerges: in the near-forward region ($x'=0.07$) where the magnitude of the amplitudes is largest, the difference between $|T_{\text{Lab}}|^2$ and $|T_{\text{CM}}|^2$ remains modest, generally bounded within $10\%$; contrastively in the non-forward region ($x'=0.86$), the difference becomes significantly larger and $|T_{\text{CM}}|^2$ exceeds $|T_{\text{Lab}}|^2$ by $\sim 40$-$150\%$ at $P=1$-$1.5$\,GeV. This pattern suggests that the CM-frame approximation is most reliable in the forward region where the Born term with small momentum transfer ($\vec{q'}\approx \vec q$ in Eq.~(\ref{momentum-space-potential})) dominates, but as the scattering angle widens (the non-forward region), the Born term is suppressed and the amplitude is dominated by more complex $P$-dependent structures through the two-particle propagator in Eq.~(\ref{T_matrix_general}).

What directly enters the evaluation of physical observables is the relativistically normalized scattering amplitude $\mathcal{M}$, which differs from the $T$-matrix by a combination of the incoming and outgoing particles' energies,
\begin{align}\label{eq:M_and_T_relation}
	\mathcal{M} = -\sqrt{(2\omega_Q) (2\omega_q)(2\omega'_Q) (2\omega'_q) } \; T\equiv-R_ET.
\end{align}
The kinematic factor $R_E$ acts as a massive amplifier of the frame-induced discrepancies in particular in the forward region, where $|\mathcal{M}_{\text{Lab}}|^2$ now reversely overtakes $|\mathcal{M}_{\text{CM}}|^2$ by $\sim 100\%$ already at $P=1.5$\,GeV, as shown in Fig.~\ref{fig:Lab-vs-CM-amplitudes-comparison}(b). Contrastively in the non-forward region, the relative difference between two amplitudes does not change much by $R_E$ when transitioning from $|T|^2$ to $|\mathcal{M}|^2$.

To illustrate the frame dependence of the scattering amplitude in a wider kinematic region relevant for the light quark momentum distributions in the QGP, we switch to another set of five variables ($p_Q,p_q,x,x',\Phi$) that equivalently dictates the scattering configurations and plot the corresponding $|\mathcal{M}|^2$ in the ($p_q,x'$) plane by fixing $p_Q=0.5$\,GeV, $x=-0.45$ and $\Phi=0$, as shown in Fig.~\ref{fig:Lab-vs-CM-amplitudes-comparison}(c) and \ref{fig:Lab-vs-CM-amplitudes-comparison}(d). In both frames, prominent peak structures emerge in the forward regions, yet exhibiting distinctly different profiles and magnitudes. For $|\mathcal{M}_{\text{Lab}}|^2$, the forward peak is localized as a sharp ridge along the $x'\sim x$ kinematic alignment. In contrast, $|\mathcal{M}_{\text{CM}}|^2$ exhibits two displaced peaks (both mapping to the forward region $\theta_{\rm CM} \sim 0$ in the CM frame), reflecting the distorted projection of the amplitude onto laboratory-frame variables by the CM-frame approximation. While the sign and size of the relative difference between these two amplitudes vary across phase space regions, we've found that this distortion persists. The striking contrast of these forward structures thus provides an intuitive measure of the frame-dependent discrepancies embedded in the in-medium non-perturbative amplitudes.

%%%%%%%%%%%%%%%%%%%%%%%%%%%%%%%%%%%%%%%%%%%%%%%%%%%%%%%%%%%%%%%%
\section{Laboratory-frame effects on heavy-quark drag}
\label{sec_lab-vs-cm-frame-drag}
%%%%%%%%%%%%%%%%%%%%%%%%%%%%%%%%%%%%%%%%%%%%%%%%%%%%%%%%%%%%%%%%
We now quantify the phenomenological impact of the laboratory-frame $T$-matrix by computing the heavy quark drag coefficient, which governs the kinetic relaxation of a heavy quark propagating through the QGP~\cite{Rapp:2009my}. The drag coefficient is obtained by integrating the momentum-transfer-weighted heayv-light collision kernel over the relativistic phase space~\cite{Svetitsky:1987gq}

\begin{align}
	\label{eq:drag_coeff_full}
	&A(p_Q) = \frac{1}{2\omega_Q} \int\frac{d^{3} p'_{Q}}{(2\pi)^{3}2\omega'_{Q}}\int\frac{d^{3} p'_{q}}{(2\pi)^{3}2\omega'_q}\int\frac{d^{3} p_{q}}{(2\pi)^{3}2\omega_q} \nonumber \\
	&\times (2\pi)^{4}\delta(\omega'_{Q}+\omega'_{q}-\omega_{Q}-\omega_{q})\delta^{3}(\vec{p'}_{Q}+\vec{p'}_{q}-\vec{p}_{q}-\vec{p}_{Q})\nonumber \\
	&\times |\mathcal{M}|^{2}(1-n_F(\omega'_q))n_F(\omega_q)\left(1- \frac{\vec{p'}_{Q}\cdot \vec{p}_{Q}}{|\vec{p}_{Q}|^2}\right),
\end{align}
where $n_F$ is the Fermi distribution for the incoming/outgoing light anti-quark. The scattering amplitude in Eq.~(\ref{eq:drag_coeff_full}), as previously emphasized, should be taken as the one evaluated in the laboratory frame. Yet in existing $T$-matrix approach to the heavy quark thermal relaxation, the CM-frame amplitude $|\mathcal{M}_{\text{CM}}|^2$ has long been employed~\cite{vanHees:2007me,Riek:2010fk,Liu:2018syc,Tang:2023tkm}, under the assumption that the in-medium amplitude is independent of $P$. The laboratory-frame formalism developed here allows to test this approximation and quantify the pertinent uncertainties. The evaluation of Eq.~(\ref{eq:drag_coeff_full}) then proceeds by expressing the heavy and light quark momenta in terms of total and relative variables and matching the collision kernel to the on-shell $|\mathcal{M}_{\text{Lab}}|^2$ upon integrating out the 4-momentum conserving $\delta$-functions.

\begin{figure}[htbp]
	\centering
	\begin{minipage}[b]{0.492\linewidth}
		\centering
		\includegraphics[width=\linewidth]{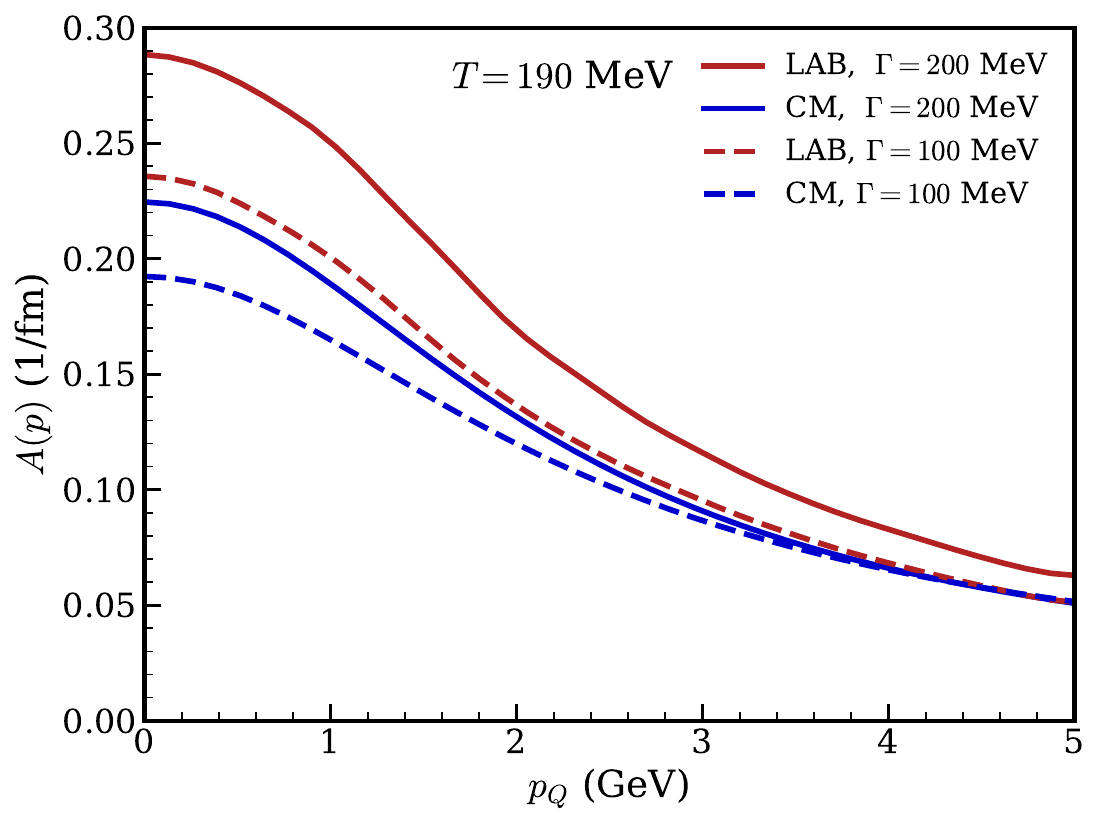}
	\end{minipage}
	\hfill
	\begin{minipage}[b]{0.492\linewidth}
		\centering
		\includegraphics[width=\linewidth]{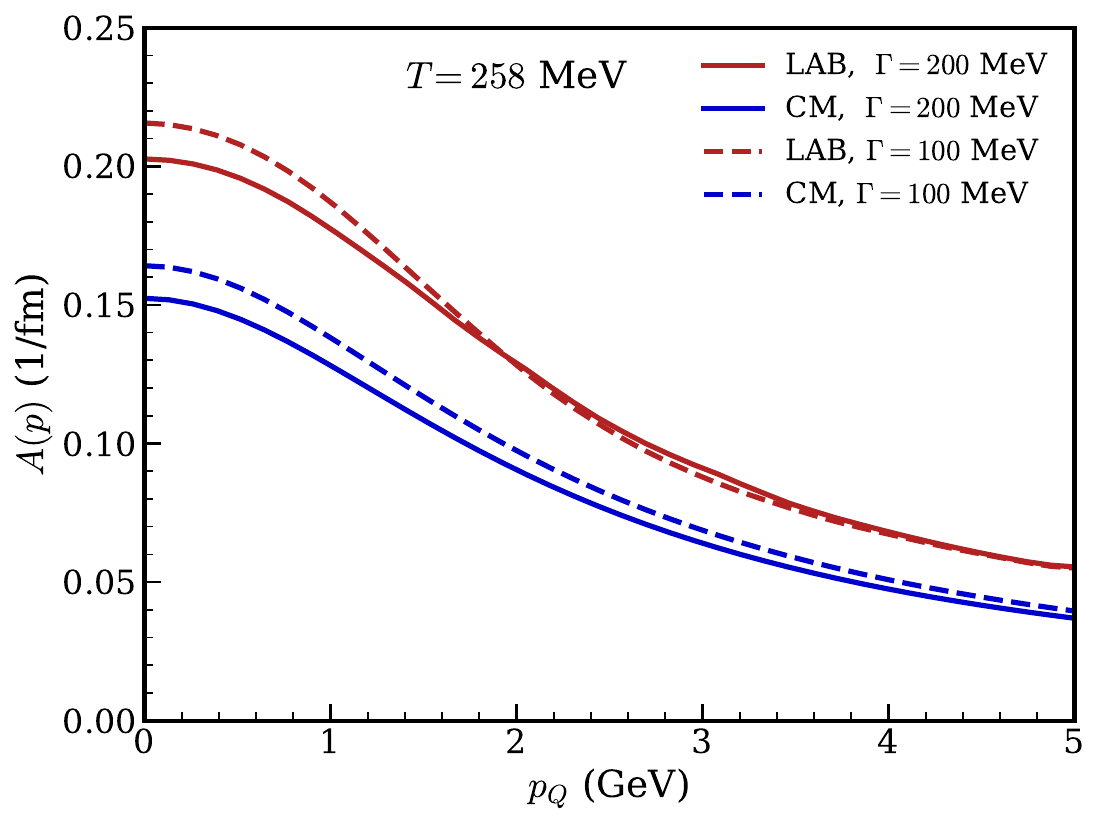}
	\end{minipage}
	\caption{Charm quark drag coefficients computed using the laboratory-frame $T$-matrix (red) in comparison with the CM-frame results (blue) at $T=190$\,MeV (left) and $T=258$ MeV (right).}
	\label{drag-Lab-vs-CM}
\end{figure}

The resulting heavy quark drag coefficients are displayed and compared in Fig.~\ref{drag-Lab-vs-CM}. While both calculations exhibit a $\sim 20$-30\% spread when the quark widths change from $\Gamma_{\rm Q,q}=100$ to $200$\,MeV, the laboratory-frame calculation yields a significant enhancement relative to the CM-frame baseline. Specifically, at the lower temperature $T = 190$\,MeV (left panel), the laboratory-frame result is $\sim 25$-$30$\% larger than the CM-frame counterpart at low heavy quark momenta $p_Q\leq 1$\,GeV, but the relative discrepancy gradually diminishes toward larger $p_Q$. At a higher $T=258$\,MeV (right panel), the relative difference between the two calculations grows to $\sim 30$-$40$\% and persists across almost the entire momentum range explored. This may be understood by the fact that at higher temperatures, the thermal motion of light quarks increases, naturally shifting the collision phase space toward higher pair-momentum $P$; concurrently, the increased weakening of the potential reduces the relative strength of the $P$-independent Born term in Eq.~(\ref{T_matrix_general}), rendering the $P$-dependent, non-perturbative integral term more prominent. Furthermore, the drag coefficient exhibits a quantitatively similar temperature evolution in both frames: increasing the temperature from $T=190$ to $258$\,MeV results in a moderate reduction (of the order 10\%-30\%, depending on the quark width) in the drag strength at vanishing heavy-quark momentum. This suggests that the primary effect of the laboratory-frame $P$-dependence is a systematic enhancement of the magnitude of the drag coefficient, rather than a modification of its temperature dependence.

%%%%%%%%%%%%%%%%%%%%%%%%%%%%%%%%%%%%%%%%%%%%%%%%%%%%%%%%%%%%%%%%
%\section{Summary \& outlook}
\section{Summary}
\label{sec_summary}
%%%%%%%%%%%%%%%%%%%%%%%%%%%%%%%%%%%%%%%%%%%%%%%%%%%%%%%%%%%%%%%%
In summary, we have addressed a long-standing but previously unquantified theoretical uncertainty in the computation of non-perturbative, in-medium scattering amplitudes: the mismatch between the CM frame, where the $T$-matrix equation is conventionally solved, and the laboratory frame, where physical observables must be constructed. We developed a practical framework to solve the in-medium two-body $T$-matrix directly in the laboratory frame, retaining the full dependence on the pair-momentum $\vec P$ and the complete scattering geometry. The key technical advance is the exploitation of the residual azimuthal symmetry about the $\vec P$-axis, which allows the high-dimensional laboratory-frame scattering equation to be decomposed into a set of independent, uncoupled integral equations for the azimuthal components $T_m$, enabling a numerically tractable implementation without additional approximation. 

Applying this framework to heavy-light quark scattering in the QGP, we demonstrated that the resulting laboratory-frame scattering amplitudes differ significantly from their CM-frame counterparts. In particular a massive distortion in the conventional CM frame shortcuts was uncovered, where forward-scattering intensity is unphysically displaced. When these amplitudes are folded into the thermal phase-space average to compute the charm quark drag coefficient, the laboratory-frame result exceeds the CM-frame baseline by 25-40\% at low momenta and moderate temperatures, while simultaneously predicting a similar temperature dependence. These findings establish that the pair-momentum dependence of the non-perturabtive, in-medium scattering amplitude - intrinsically neglected in conventional CM-frame based calculations - constitutes a significant and previously unaccounted source of theoretical uncertainty, and its proper treatment yields a correction that is physically consequential.

The framework established here opens up several avenues for extensions and applications. An immediate step is to extend the laboratory-frame calculation to include heavy quark-gluon scattering.
%, completing the full non-perturbative interaction portrait for heavy quarks in the medium. 
Furthermore, a fully self-consistent calculation~\cite{Liu:2017qah} in which the single-particle self-energies and spectral functions are determined iteratively alongside the laboratory-frame $T$-matrix should be pursued, which would eliminate the constant widths scenario and yield a genuinely self-contained, conserving description. The transport coefficients thus obtained should then be implemented into Langevin-type transport simulations to assess the pertinent impact on experimental observables~\cite{He:2022ywp}, which is essential for reliable inferences of the transport properties of the QGP from precision heavy flavor data in relativistic heavy-ion collisions~\cite{ALICE:2022wpn,ALICE:2022wwr,CMS:2024irj}. Finally, because the explicit breaking of Lorentz invariance by a thermal or dense background is a universal feature of many-body systems, the present framework can be adapted to other non-perturbative in-medium scattering problems, {\it e.g.} reactions in finite-density nuclear matter, where the laboratory-frame $T$-matrix similarly dictates the true physical observables.

\acknowledgments This work was supported by NSFC grant 12475141.

\end{document}